\documentclass[aps,prl,preprint,groupedaddress,onecolumn,longbibliography]{revtex4-1}
\usepackage{graphicx,amsmath,amssymb}
\usepackage[percent]{overpic}
\usepackage{color}
\usepackage{bm}
\usepackage[normalem]{ulem}
\newcommand{\sjt}[1]{{\color{black}#1}}

\begin{document}

\title{Collective vibrations of confined levitating droplets}

\author{S. J. Thomson}
\email{Corresponding author: thomsons@mit.edu}

\affiliation{Department of Mathematics, Massachusetts Institute of Technology, Cambridge, MA 02139}
\author{M. M. P. Couchman}
\affiliation{Department of Mathematics, Massachusetts Institute of Technology, Cambridge, MA 02139}

\author{J. W. M. Bush}
\affiliation{Department of Mathematics, Massachusetts Institute of Technology, Cambridge, MA 02139}

\date{\today}

\begin{abstract}
We report a new type of fluid-based driven dissipative oscillator system consisting of a lattice of millimetric fluid droplets bouncing on a vertically vibrating liquid bath and bound within an annular ring. We characterize the system behavior as it is energized through a progressive increase in the bath's vibrational acceleration. Depending on the number of drops, the onset of motion of the lattice may take the form of either out-of-phase oscillations or a striking solitary wave-like instability. Theoretical modeling demonstrates that these behaviors may be attributed to different bifurcations at the onset of instability. The results presented here demonstrate the potential and utility of the walking droplet system as a platform for investigating wave-mediated, inertial, non-equilibrium particle dynamics at the macroscale.
\end{abstract}


\maketitle

\emph{Introduction}---Emergent wave propagation in fluid-based, many-body systems, both passive and active, is an area of burgeoning interest, having been observed in a wide range of settings including bacterial suspensions \cite{dunkel2013fluid,creppy2016symmetry,wioland2016directed,theillard2017geometric}; colloidal fluids composed of synthetic micro-rollers and spinners \cite{bricard2015emergent,geyer2018sounds, soni2019odd}; and crystals of driven microfluidic water droplets \cite{beatus2006phonons,janssen2012collective,schiller2015collective,tsang2018activity}. Elsewhere, theoretical models of one-dimensional driven dissipative lattices exhibit instabilities in the form of solitary waves, as well as unidirectional motion and out-of-phase (optical) oscillations \cite{ebeling2000nonlinear,makarov2000soliton,dunkel2002coherent,chetverikov2003phase,chetverikov2006dissipative,chetverikov2018dissipative,chetverikov2019noise}, prompting experimental realizations in the form of active electronic circuits \cite{hirota1973theoretical,singer1999circuit,makarov2001dissipative, nekorkin2012synergetic,kotwal2019active}. We here introduce a new, robust, mechanical analogue of an active nonlinear lattice, comprised of quasi-one-dimensional assemblies of self-propelled fluid droplets.

When introduced onto the surface of a vibrating liquid bath, millimetric droplets have been shown to bounce or ``walk'' across the interface by means of self-propulsion through a resonant interaction with their own wave field \cite{couder2005bouncing,couder2005dynamical,protiere2006particle,bush2015pilot}. In the bouncing state, the net wave force exerted on the drop by the bath supports its weight, enabling it to levitate above the bath surface, precluding coalescence. Above a critical vibrational acceleration, the droplet becomes unstable to small lateral perturbations. When the resulting propulsive wave force overcomes the stabilising effects of dissipation, the drop begins to walk \cite{molavcek2013dropsb}. \sjt{Droplets brought into close proximity interact and may become coupled through their common wave field, a phenomenon which has prompted several investigations into the dynamics and stability of droplet pairs \cite{protiere2008exotic,eddi2012level,borghesi2014interaction, oza2017orbiting, arbelaiz2018promenading,galeano2018ratcheting,couchman2019bouncing} and two-dimensional lattices \cite{lieber2007self,eddi2009archimedean,eddi2011oscillating}.} We here consider the collective dynamics of a quasi-one-dimensional lattice of coupled droplets (Figure 1(a)), systematically characterizing the behaviour of the system as the vibrational acceleration of the bath is increased progressively. We demonstrate that this driven dissipative oscillator exhibits two distinctive emergent features, specifically out-of-phase oscillations and solitary wave propagation, \sjt{neither of which have hitherto been reported in the bouncing-droplet literature.  Using the theoretical model and stability analysis presented in \cite{thomson2020collective}, we rationalize a novel bifurcation structure responsible for these distinct dynamical states. The results presented here demonstrate the potential and utility of the walking droplet system as a platform for investigating wave-mediated, inertial, non-equilibrium particle dynamics at the macroscale \cite{bechinger2016active,klotsa2019above}.}

\begin{figure}[htp]
    \centering
    \includegraphics[width=0.9\textwidth]{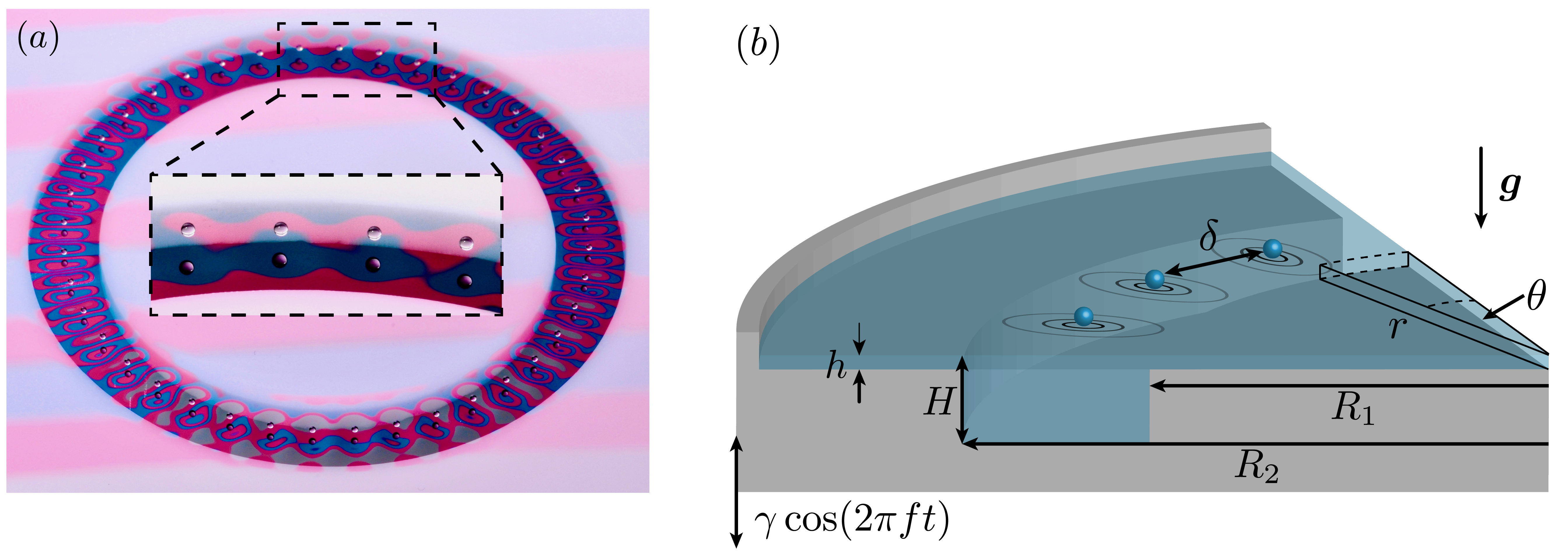}
    \caption{(a) Oblique perspective of a lattice of 40 equispaced droplets of silicone oil confined to an annular channel. (b)~Schematic of the channel geometry with inner radius~$R_{1} = 24$ mm and outer radius $R_{2} = 31$ mm. A shallow layer of depth $h = 0.56\ \text{mm}$ acts as a damper so that wave motion is largely confined to the channel where the fluid depth $H = 5.12\ \text{mm}$. The entire assembly is vibrated vertically with maximum acceleration $\gamma$ and frequency $f$.}
    \label{fig:exp_fig}
\end{figure}

\begin{figure*}
\centering
\includegraphics[width=\textwidth]{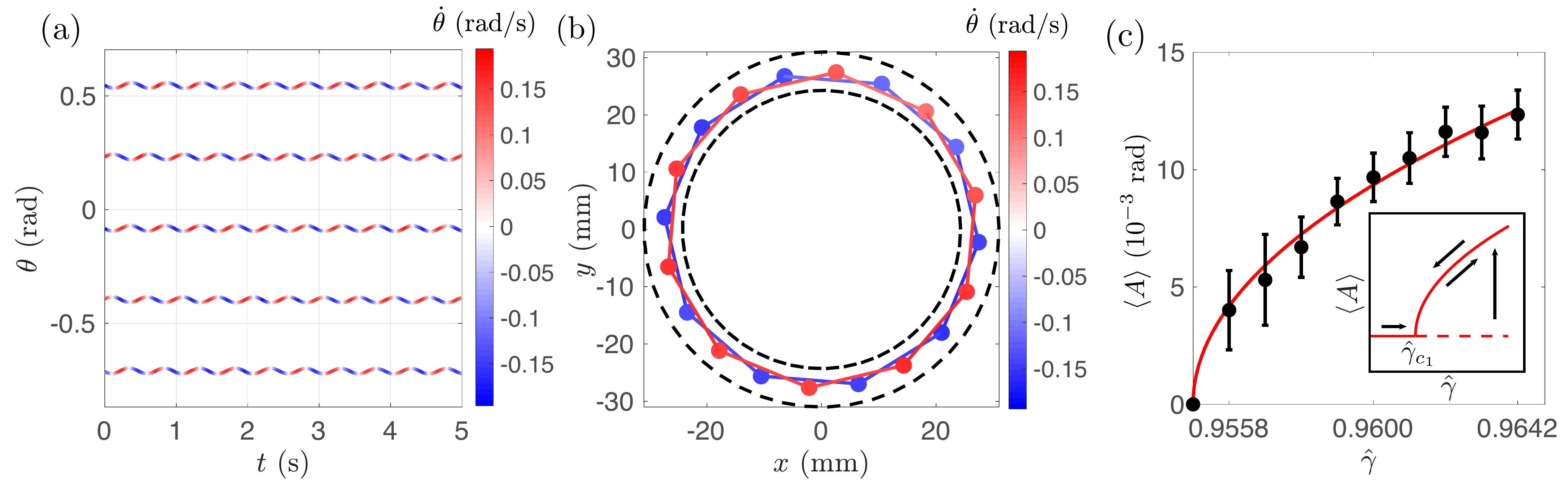}
\caption{$N = 20$ droplets. (a) Evolution of the polar angle $\theta$ of six adjacent droplets undergoing out-of-phase oscillations with frequency $F = 2$ Hz at vibrational acceleration $\hat{\gamma} = 0.962$. The trajectories are colored according to the droplets' instantaneous angular velocity $\dot{\theta}$. Note that neighboring droplets have equal and opposite $\dot{\theta}$ for all time $t$. (b) A snapshot of the polar positions of each droplet in the lattice at a time when each droplet attains its maximum angular speed of $|\dot{\theta}| = 0.16$ rad s$^{-1}$. The value of $\hat{\gamma}$ is the same as in (a). The solid lines emphasize the net out-of-phase oscillation of two decahedral sub-lattices and the dashed lines delineate the channel geometry. (c) Plot of the mean oscillation amplitude $\langle A \rangle$ as a function of $\hat{\gamma}$ shows a monotonic increase of $\langle A \rangle$ beyond a critical value of $\hat{\gamma}_{c_{1}} = 0.956$. The error bars represent the standard deviation of the mean. The red curve is a square-root-fit of the data. Inset: a schematic of the reversible supercritical bifurcation path underpinning the observed instability. The solid red line indicates the stable branch consisting of static then oscillating droplets; the dashed red line is unstable.}
\label{fig:fig2}
\end{figure*}
\emph{Experiments}---Our experimental set-up is based on that described in~\cite{harris2015generating}. An annulus of inner radius $R_{1} = 24$ mm and outer radius $R_{2} = 31$ mm was placed in a stainless steel bath and mounted on an electromagnetic shaker. \sjt{The dimensions of the annulus were tuned to accomodate 40 droplets. When arranged into a quasi-one-dimensional lattice of sychronized bouncers, these 40 drops occupy their preferred inter-droplet spacing $\delta\approx 4.32$ mm, as is dictated by the superposition of wave fields generated by the individual droplets.} The bath was filled with silicone oil of density $\rho =~950\ \text{kg m}^{-3}$, viscosity $\nu = 20$ cSt, and surface tension $\sigma = 20.6\ \text{mN}\ \text{m}^{-1}$, to a depth $H = 5.12\ \text{mm}$ inside the annulus and $h =~0.56\ \text{mm}$ outside. The shallow region acts as a wave damper, ensuring that all droplet motion is confined to the channel, a technique employed elsewhere \cite{filoux2015strings,filoux2017walking}. A schematic of our experiment is shown in Figure \ref{fig:exp_fig}(b). The set-up was vibrated vertically with acceleration~$a(t) = \gamma\cos(2\pi f t)$,~where $f$ and $\gamma$ denote the frequency and maximum acceleration respectively. The vibration of the bath was monitored using two accelerometers and a closed-loop feedback ensured that $\gamma$ was kept constant to within $\pm 0.002g$ . The frequency $f$ was fixed at 80 Hz. We note that a change in $f$ can dramatically alter the system properties, influencing not only the wavelength of Faraday waves excited on the bath surface, and so the preferred interdroplet spacing, but the stability characteristics of the droplet lattice. For the chosen control parameters and geometry, the Faraday threshold \cite{bush2015pilot} in the channel was found to be $\gamma_{F} = 4.72 \pm 0.05g$. Over time, $\gamma_{F}$ may slowly decrease due to the liquid temperature increasing and the concomitant decrease in viscosity \cite{harris2014droplets}. To minimize this effect, the bath was vibrated for 1 hour at $\gamma > \gamma_{F}$ before the start of the experiment and $\gamma_{F}$ was measured both before and after the experiment, its drift monitored. To eliminate the influence of air currents, the bath was surrounded by a transparent acrylic box \cite{pucci2018walking}.

Droplets of diameter $D = 0.72 \pm 0.01$ mm were generated using a piezoelectric drop generator and placed into the channel with the aid of a prewetted slide \cite{harris2015low}. The presence of the channel and surrounding shallow layer largely restricts the droplet motion along the azimuthal direction $\theta$, however the droplets are free to move slightly in the radial direction $r$. The droplets were arranged into an equispaced lattice at $\gamma = \gamma_{B} = 4g$, bouncing in period synchrony with period $T = 2/f$ \footnote[1]{See Supplemental Material at \text{xxx} for access to all experimental videos and supporting materials referenced in this paper}, resonating with their subcritical Faraday waves \cite{bush2015pilot,molavcek2013drops}. At equilibrium, the droplets sit at a radius of $R_{m} = 27.5$ mm and are separated by an angle $\Delta\theta = 2\pi/N$. In what follows we describe the outcomes of two sets of experiments: one involving a lattice of $N = 20$ droplets and the other $N = 40$ droplets, \sjt{with inter-droplet spacings of approximately $2\delta$ and $\delta$, respectively. We note that the inter-droplet spacing cannot be varied continuously; rather, it is selected by the sub-threshold Faraday waves produced by the droplets. With our fixed annular geometry, we are thus unable to study the dynamics of arbitrary $N$-droplet lattices. We further note that removal of a droplet from both the $N = 20$ and $N = 40$ configurations resulted in a lattice vacancy, rather than an increase in the inter-droplet spacing. The addition of a droplet caused the lattice to buckle in the radial direction, $r$, destroying its quasi-one-dimensional structure.} The spatio-temporal coupling of the droplets is mediated by the wave field. However we note that a novelty in our system is that this effective potential is not fixed in space with respect to the droplet; rather, it is dynamically generated by, and continuously evolves with, the droplet motion. 

\begin{figure*}
    \centering
    \includegraphics[width=\textwidth]{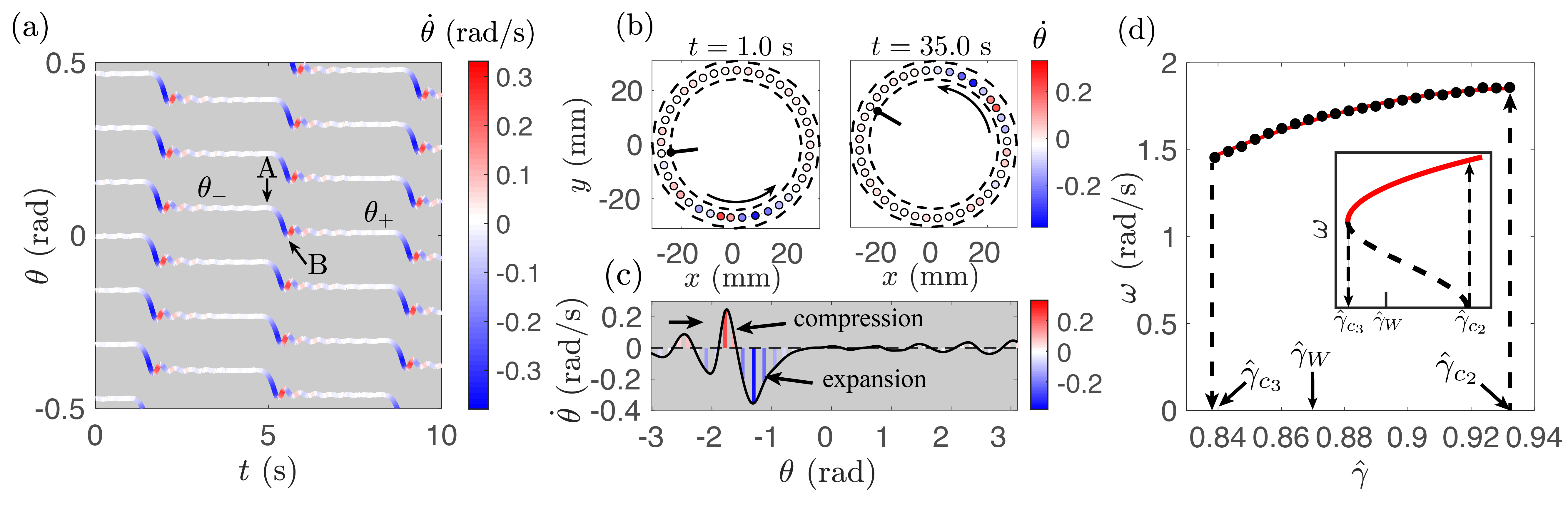}
    \caption{ $N = 40$ droplets. (a) Evolution of the angular position $\theta$ of individual droplets in the lattice in the solitary-wave regime at vibrational acceleration $\hat{\gamma} = 0.890$. The trajectories are colored according to the droplets' instantaneous angular velocity $\dot{\theta}$. Prior to the arrival of the wave at event A, a droplet is stationary at $\theta_{-} \approx 0.08$ rad. In the interval AB, it is pulled in the direction of decreasing $\theta$ over a time of order $0.5$ s, reaching a maximum angular speed of $|\dot{\theta}| \approx 0.38$ rad s$^{-1}$. At event B, its angular speed returns to zero and thereafter the droplet settles down to $\theta_{+} \approx 0$ rad \emph{via} underdamped oscillations. (b) Polar positions of the droplets in the lattice at $t = 1.0$ s and $t = 35.0$ s. Each droplet is colored according to its instantaneous angular velocity. The value of $\hat{\gamma}$ is the same as that in (a). The black marker highlights the overall clockwise rotation of the lattice and black arrows show the direction of wave propagation. See [47] for a color-coded animation of the polar droplet positions. 
 (d) Droplet instantaneous angular velocity $\dot{\theta}$ as a function of droplet position at $t = 1$ s. We fit a cubic spline interpolant through the maximum of each spike, revealing a well-defined wave packet propagating in an anti-clockwise direction through the lattice. (e) Plot of $\omega(\hat{\gamma})$ (black dots). We find $\omega$~varies monotonically with $\hat{\gamma}$ but drops to zero at $\hat{\gamma}_{c_{3}} = 0.839$. The best fit red curve emphasizes the nonlinearity of $\omega(\hat{\gamma})$. The maximum error in $\omega$ was found to be $\pm 0.01$ rad s$^{-1}$. Inset: a schematic sketch of the subcritical bifurcation diagram underpinning the solitary wave instability. The upper solid red line (symbolic of the best fit line in the main plot) denotes the stable solitary wave branch, the dashed black curve a hypothetical unstable branch.}
    \label{fig:waves}
\end{figure*}

The acceleration of the bath was incremented in steps of $0.01g$ from $\gamma_{B}$ until the stationary bouncing state destabilized. Thereafter, $\gamma$ was varied in steps of $\Delta\gamma = 0.005g$ to increase the resolution of our data. Droplet positions were tracked using a CCD camera mounted directly above the bath, recording at 20 frames-per-second, and then processed using an in-house droplet-tracking algorithm in MATLAB. Henceforth, we characterize the proximity to the Faraday threshold in terms of the dimensionless vibrational acceleration~$\hat{\gamma} =~\gamma/\gamma_{F}$. For reference,~$\hat{\gamma}_{B} = 0.847$ and the walking instability threshold of a single droplet $\hat{\gamma}_{W} = 0.873 \pm 0.02$.

With $N = 20$ droplets, the onset of instability was manifest as small azimuthal oscillations of the droplets about their equilibrium position. Eventually, synchrony between the droplets emerged in the form of out-of-phase oscillations (see \footnote[1]{} for an accompanying video). The motion of a subset of six droplet trajectories is plotted in Figure 2(a) for $\hat{\gamma} = 0.962$. Each droplet oscillates half a period out-of-phase with its neighbour with frequency $F = 2$ Hz. The net result of the instability is the out-of-phase oscillation of two decahedral sub-lattices (see Figure 2(b)). We determine the amplitude of oscillation as a function of $\hat{\gamma}$ by taking the mean oscillation amplitude $\langle A \rangle$ of all droplets in the lattice. As shown in Figure 2(c), $\langle A \rangle$ increases monotonically from zero at $\hat{\gamma}_{c_{1}} = 0.956 \pm 0.02$ \sjt{to its maximum at $\hat{\gamma} = 0.9642$, where relatively large droplet oscillations cause the periodic oscillatory state to destabilise, leading to an effective melting of the lattice}. We note that the error in $\hat{\gamma}_{c_{1}}$ is due to a small hysteresis that arises when the instability threshold is approached from below or above \cite{molavcek2013dropsb,wind2013exotic}. The foregoing observations point to a reversible supercritical bifurcation (see inset of Figure 2(c)) underpinning the observed instability.

With a lattice of $N = 40$ droplets, multiple realizations of the same experiment revealed that the \sjt{stable, out-of-phase oscillations apparent for $N = 20$ are never observed; instead, the lattice initially destabilizes to irregular azimuthal oscillations}. After a short transient, one droplet in the lattice eventually receives a large amplitude kick in the $\theta$ direction, prompting a markedly different dynamics, namely the spontaneous excitation of a solitary wave that propagates around the ring indefinitely (see \footnote[1]{} for an accompanying video of this transition). The direction of propagation was either clockwise or counter-clockwise with equal probability. The dramatic transition to the solitary wave regime occurred at $\hat{\gamma}_{c_{2}} = 0.932\pm 0.02$ and was observed in some instances to trigger a second wave following in the wake of the first. However it was found that this second wave could be suppressed by slightly reducing the acceleration. The experiments and accompanying data described here are for $\hat{\gamma}\in (0.839,0.932)$, in which range a single solitary wave propagates around the lattice. 

The wave progresses around the ring as each droplet successively undergoes the motion shown in Figure \ref{fig:waves}(a). Prior to the arrival of the wave an individual droplet is essentially stationary before being pulled sharply in the direction of decreasing $\theta$. It then receives a restoring force in the opposite direction before settling down to a new \sjt{static} equilibrium position \emph{via} underdamped oscillations. \sjt{We note that neighboring droplets do not enter into out-of-phase oscillations following the passage of the solitary wave.} The initial jump in the droplet position gives rise to the curious feature that, while there is evidently a disturbance propagating in a counter-clockwise direction, the net displacement of the entire droplet lattice is clockwise. The wave field generated by each droplet has a spatial extent exceeding the equilibrium droplet spacing; hence, each droplet can influence more than its nearest neighbor. As shown in Figures \ref{fig:waves}(b) and (c), the current position of the wave is spread over a core of approximately 7 droplets undergoing different stages of the motion described in Figure \ref{fig:waves}(a).
Following the passage of the wave, the droplets oscillate back and forth and eventually recover their uniform initial spacing. 

After increasing the acceleration so as to excite a solitary wave, we computed $\omega$, the angular frequency of the wave, as a function of $\hat{\gamma}$ 
in the interval $\hat{\gamma}\in (0.839,0.932)$, which revealed a monotonic relationship between the two (see Figure \ref{fig:waves}(d)). We note that, in contrast to the $N = 20$ case, this interval contains the walking threshold of a single droplet $\hat{\gamma}_{W} = 0.873$. Determining $\omega(\hat{\gamma})$ revealed two further differences with the case of $N = 20$ droplets. First, we find that the solitary wave can propagate without decay far below the initial instability threshold of the stationary lattice $\hat{\gamma}_{c_{2}}$, specifically for $\hat{\gamma} < \hat{\gamma}_{W}$. Second, we find that there is a critical value 
$\hat{\gamma}_{c_{3}} = 0.839$ at which $\omega$ jumps \emph{discontinuously} to zero from a finite value (for a video of this transition, see \footnote[1]{}). 
In the parlance of dynamical systems, the point $\hat{\gamma}_{c_{3}}$ is reminscent of a saddle node or blue-sky bifurcation \cite{strogatz2018nonlinear}. We thus deduce the coexistence of two qualitatively different stable states of the system for $\hat{\gamma}\in (0.839,0.932)$: the static equilibrium configuration of the bouncing droplets and a periodic state consisting of the solitary wave. The foregoing observations suggest that the system has undergone a subcritical bifurcation. A sketch of the associated hysteresis path is plotted in the inset of Figure \ref{fig:waves}(d) and is to be contrasted with that of Figure \ref{fig:fig2}(c). 

\sjt{\emph{Theoretical modelling}}---We now summarize a mathematical model and accompanying stability analysis, presented in \cite{thomson2020collective}, aimed at rationalizing qualitatively the different bifurcations that can arise as the number of droplets, $N$, varies. The model presented in \cite{thomson2020collective} considers $N$ equispaced droplets of equal mass which bounce in periodic synchrony, confined to a circle of constant radius $R$. The arc-length position of each drop $x_{n}$ evolves according to the stroboscopic model derived by Oza \emph{et al}.\ \cite{oza2013trajectory}, which describes the time-averaged motion of the droplet in terms of a balance between inertia, drag, and the propulsive wave force enacted on each droplet as it lands on the sloping crest of its local wave field. As shown in \cite{thomson2020collective}, the radially symmetric, $N$-droplet analogue of the stroboscopic model, valid below and near to the point of instability, in dimensionless form reads
\begin{subequations}
\label{eqn:model}
\begin{equation}
\label{eqn:xn}
\ddot{x}_{n} + \dot{x}_{n} = -\mathcal{H}_{x}(x_{n},t),
\end{equation}
where the wave field
\begin{equation}
\label{eqn:h}
\mathcal{H} = \sum_{m = 1}^{N}\int_{-\infty}^{t} \mathcal{K}(x - x_{m}(s))\text{e}^{-(t - s)/M}\ \text{d}s.
\end{equation}
\end{subequations}
The sum over $N$ in equation \eqref{eqn:h} represents the superposition of the wave fields generated by each droplet in the lattice, while the integral represents the physical fact that the trajectory of each droplet depends on its entire history prior to the current time $t$, endowing the system with memory \cite{eddi2011information,turton2018review}.

The model is closed by selecting a particular form for the stroboscopic wave field kernel $\mathcal{K}$. Owing to the geometry of the experimental system and the presence of submerged topography, there is some uncertainty as to the precise form of the droplet wave field as compared to free space. The situation is complicated yet further by variations in the droplets' vertical bouncing phase and memory-dependent changes to the decay length of the emitted Faraday waves \cite{tadrist2018faraday,couchman2019bouncing}. Following \cite{thomson2020collective}, we thus consider the generic pilot-wave model $\mathcal{K}(x) = f(2R\sin(x/2R))$, which is derived by taking a circular cut of radius $R$ of the axisymmetric wave field $f(r) = \mathcal{A}J_{0}(2\pi r)\text{sech}(r/l)$, centered a distance $R$ from the origin (further details are provided in the Supplementary Material). The parameter $l$ is the dimensionless decay length of the waves relative to the Faraday wavelength and $\mathcal{A}$ is the dimensionless wave amplitude. Unless otherwise stated, we take $l = 1.6$, $\mathcal{A} = 0.1$, and $R = 5.4$, based on typical experimental values \cite{eddi2011information,molavcek2013drops}. The remaining parameter $M = \mu(1 - \hat{\gamma})^{-1}$ is the memory parameter which encodes the exponential decay time of the waves, where the constant $\mu = T_{d}D/m = 0.14$ is composed of the temporal decay time of Faraday waves $T_{d}$, the drag coefficient $D$, and the droplet mass $m$ \cite{eddi2011information,molavcek2013drops,oza2013trajectory}. Thus, $M\rightarrow\infty$ as $\gamma\rightarrow\gamma_{F}$ and larger $M$ corresponds to past dynamics playing a more prominent role.

Before we proceed, a few comments are in order. First, the chosen form of the wave field kernel, $\mathcal{K}$, will inevitably preclude a quantitative match between theoretical and experimental predictions of the instability thresholds. Further, the wave amplitude $\mathcal{A}$ depends in reality on the vertical bouncing phase of the droplets \cite{couchman2019bouncing}, and thus is another source of discrepancy between theoretical and experimental predictions of the instability thresholds. However, it was found that varying $\mathcal{A}$ acted only to shift the numerical value of the predicted instability threshold, with the bifurcation structure remaining virtually unchanged. This suggests that the stability of the lattice has more to do with the geometry of the system and the resulting global wave field $\mathcal{H}$, and less on the details of the specific parameters used.  Finally, the model \eqref{eqn:model} prohibits radial motion of the droplets, and thus we do not capture buckling of the lattice, nor the fully nonlinear dynamics of the solitary wave, where radial motion of the droplets can be significant. A detailed study of the specific fluid system explored experimentally is subject to future work.
\begin{figure}[h]
\centering
\includegraphics[width=0.75\textwidth]{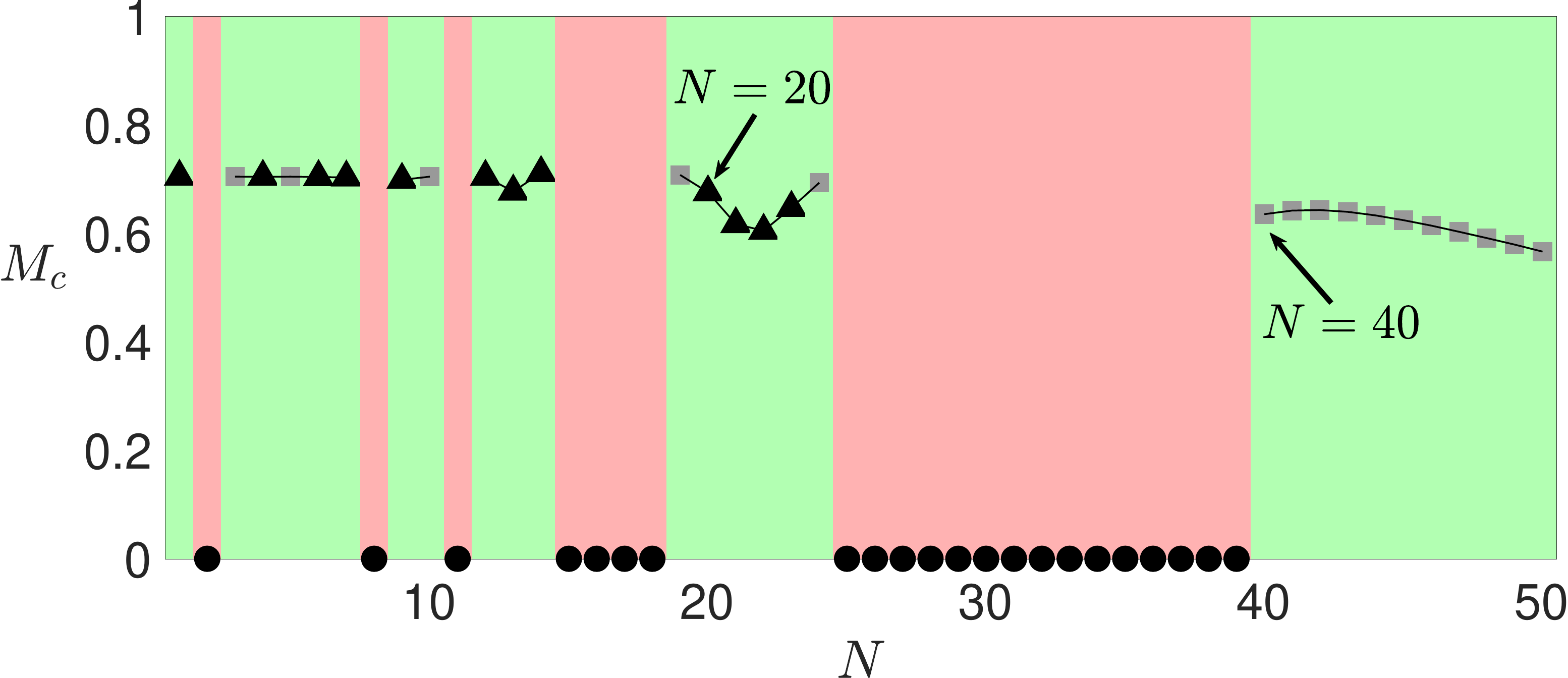}
\caption{Stability threshold $M_{c}$ versus droplet number $N$ predicted by the model \eqref{eqn:model}, disallowing for radial motion. Triangles mark supercritical bifurcations $(\Re(\beta_{2}) > 0)$ and squares mark subcritical bifurcations $(\Re(\beta_{2}) < 0)$ as predicted by Equation \eqref{eqn:stuart_landau}. Red zones and circles denote droplet configurations predicted to be unstable for all $M$, and so inaccessible in the laboratory. Green zones denote configurations predicted to be stable up to a critical value of $M$, and thus potentially accessible in the laboratory. Figure adapted, with permission, from \cite{thomson2020collective}.}
\label{fig:fig4}
\end{figure}

The stability characteristics of the lattice may be rationalized qualitatively by a weakly-nonlinear analysis of the system \eqref{eqn:model} in the vicinity of the bifurcation point $M = M_{c}$ (that is, $M - M_{c} = \varepsilon^2$, where $\varepsilon\ll 1$), corresponding to a particular value of $\hat{\gamma}$. The analysis presented in \cite{thomson2020collective} demonstrates that the complex amplitude $A$ of the perturbation to the $n$-th droplet position, $\hat{x}_{n} = A(T)\exp(\text{i}(k_{c}n\alpha + \omega_{c}t)) + \text{c.c.}$, may be described by a Stuart-Landau equation of the form
\begin{equation}
\label{eqn:stuart_landau}
\frac{\text{d}A}{\text{d}T} = \beta_{1}(N)A - \beta_{2}(N)|A|^2 A,
\end{equation}
where the droplet separation $\alpha = 2\pi/N$ and $A$ is a function of the slow time-scale $T = \varepsilon^2 t$. The critical wavenumber $k_{c}$ and angular frequency $\omega_{c}$ are determined from the linear stability of \eqref{eqn:model}. The crucial component of \eqref{eqn:stuart_landau} is the sign of $\Re(\beta_{2})$, defined in terms of the system parameters. For $\Re(\beta_{2}) > 0$ the cubic nonlinearity stabilizes linear growth, leading to a supercritical bifurcation, while $\Re(\beta_{2}) < 0$ heralds a subcritical bifurcation. 

In Figure \ref{fig:fig4} we plot the critical value of $M = M_{c}$ at which each lattice configuration destabilises given $N$, as deduced from the computations in \cite{thomson2020collective}. Droplet configurations contained within green zones of Figure \ref{fig:fig4} are stable up to a finite value of $M$ and subsequently destablise \emph{via} a Hopf bifurcation. Red zones contain configurations in this one-dimensional setting where the wavefield $h$ acts to destabilise the droplets for all $M$---as can occur if a droplet sits on a peak of the lattice wave field \cite{couchman2019bouncing}---and are thus prohibited from forming. We note that Figure \ref{fig:fig4} only documents configurations which are potentially experimentally accessible, disallowing for radial motion of the droplets. We anticipate some configurations of larger droplet density may be susceptible to radial buckling in practice, as can occur with the addition of a droplet to an otherwise stable lattice. Nevertheless, we see that the critical features of our experimental observations are captured in Figure \ref{fig:fig4}. When $N = 20$, there is a supercritical bifurcation at $M_{c} = 0.676$ ($\hat{\gamma} = 0.793$), wherein linear theory predicts $k_{c} = N/2$. Hence, we find that $\hat{x}_{n} \propto (-1)^{n}\cos(\Omega t)$, corresponding to small-amplitude, stable, out-of-phase oscillations arising beyond the instability threshold (recall Figure \ref{fig:fig2}), the angular frequency $\Omega$ being determined as part of the weakly-nonlinear analysis. A configuration of $N = 40$ droplets destabilizes to a subcritical Hopf bifurcation at $M_{c} = 0.636$ ($\hat{\gamma} = 0.780$), wherein the system approaches a distant attractor, manifest in experiments as a solitary wave (Figure \ref{fig:waves}(d)). There is a notable prevalence of subcritical bifurcations for $N\ge 40$ and tightly packed droplets, while supercritical bifurcations appear more frequently for lower droplet densities.

\emph{Discussion}---We have considered the collective vibrations of a new type of driven dissipative oscillator, which exhibits out-of-phase oscillations and solitary waves. The former is characterised by the onset of a reversible supercritical bifurcation where the amplitude of the oscillations increases with the driving acceleration. The latter subcritical bifurcation displays the coexistence of two distinct stable states, one static, the other characterized by a solitary wave whose angular frequency $\omega$ has a nonlinear dependence on $\hat{\gamma}$. The transition between these two regimes has been rationalised through a systematic stability analysis in the vicinity of the bifurcation point \cite{thomson2020collective}. A detailed description of the fully nonlinear dynamics of the system, including modelling of the fluid \cite{durey2020faraday}, is subject to future work. \sjt{There is growing interest in the walking droplet system as a platform for studying active non-equilibrium particle dynamics at the macroscale. For example, recent experiments show that collective magnetic order may arise in wave-mediated, hydrodynamic spin lattices of walking droplets \cite{saenz2018spin}. Additionally, the results presented here have prompted investigation into the wider class of instabilities available to free rings of droplets unconfined by bottom topography \cite{couchman2020rings}. In a broader context, the role of inertia differentiates the walking droplet system from prevailing active matter systems---for example, bacterial and colloidal suspensions---wherein the dynamics are typically overdamped. Indeed, that solitary-like waves are supported by the system presented here points to the existence of a wider class of self-sustaining nonlinear waves in inertial, underdamped active systems \cite{bechinger2016active,scholz2018inertial,klotsa2019above}.} Several interesting questions also arise regarding the emergence and control of solitary waves in mechanical media \cite{coste1997solitary,deng2017elastic,zhang2019programmable}, as well as prompting future investigations into connections with extant physical systems, such as zigzag transitions in low-dimensional trapped ionic crystals \cite{fishman2008structural,shimshoni2011quantum,schneider2012experimental} and emergent chimera states in oscillators subject to nonlocal spatio-temporal coupling \cite{abrams2004chimera,sethia2008clustered,martens2013chimera}.

S.J.T would like to thank Matthew Durey and Rodolfo R.\ Rosales for valuable discussions regarding the different bifurcations outlined in this paper, that prompted the careful data acquisition leading to Figures \ref{fig:fig2}(c) and \ref{fig:waves}(e). The authors would also like to thank the anonymous referees whose comments led to the improvement of this article.

The datasets used in this study are available from the corresponding author upon reasonable request. 

\bibliography{paperfinal}

\end{document}